\documentstyle[aps]{revtex}
\tightenlines

\def\be{\begin{equation}}
\def\ee{\end{equation}}
\def\ben{\begin{eqnarray}}
\def\een{\end{eqnarray}}

\title{Lorentz Invariant Superluminal Tunneling}
\author{Partha Ghose and M. K. Samal}
\date{S. N. Bose National Centre for Basic Sciences, JD/III, Salt
Lake, Kolkata 700 098, India.}

\begin{document}
\maketitle
\begin{abstract}{It is shown that superluminal optical signalling is
possible without violating Lorentz invariance and causality via
tunneling through photonic band gaps in inhomogeneous dielectrics
of a special kind.}\end{abstract}
\section{Introduction}

A number of recent experiments have reported the observation of
electromagnetic waves propagating with velocities larger than $c$
(the velocity of light in vacuum) in dispersive media \cite{Wang},
wave guides \cite{Mugai}, electronic circuits \cite{Mitchell} and
in tunneling \cite{Steinberg}. The experimenters have been quick
to point out that these observations do not necessarily contradict
the special theory of relativity and causality. These claims have
naturally generated a controversy in the literature.

In the case of dispersive propagation the claim is in apparent
contradiction with the pioneering work of Sommerfeld and Brillouin
\cite{Somm} who clearly showed the difference between group
velocity, phase velocity and signal front velocity, and
established the result that no physical signal can travel faster
than $c$ in dispersive media. However, it has recently been argued
that for physical signals that are of finite duration the
causality principle ``cause precedes effect'' is preserved despite
superluminal motion. This is because a superluminal signal
travelling backward in time can never arrive before the primary
signal is generated, thus preventing the original user changing
the transmitted signal \cite{nimtz}.

In the case of frustrated total internal reflection (FTIR) and
tunneling, the situation is quite different. It has been argued
that in such cases the wave number is imaginary, the phase is a
constant and the concept of a signal front is meaningless
\cite{nimtz}. Further, it has been pointed out that if the signal
is narrow-band limited, there is no distortion of the signal
envelope and its delay is the same as that of its centre of
gravity \cite{smith}. Since the evanescent (exponentially damped)
component of a wave does not oscillate with distance, it does not
accumulate any phase and can therefore propagate through the
evanescent region with zero (phase) delay. It has been argued that
there is empirical evidence of this in, for example, symmetrical
FTIR in which there is no time lag between the reflected and
tunneled signals \cite{nimtz}. However, it is not quite clear how
a zero phase delay necessarily implies a zero signal delay.

One source of confusion in the literature, in our opinion, is the
popular use of an analogy between the Helmholtz and Schr\"odinger
equations. Since Maxwell's equations in an inhomogeneous but
isotropic medium reduce to the Helmholtz equation for a
monochromatic wave in the scalar approximation, and the Helmholtz
and the {\it non-relativistic} Schr\"odinger equations are
formally identical, the one-dimensional process of
non-relativistic quantum mechanical tunneling has been used to
model the optical process of transmission through a barrier
\cite{nimtz}. This is obviously unsatisfactory, because (a) the
Schr\"odinger evolution used is characteristically
non-relativistic whereas the optical processes in question are
intrinsically relativistic, and (b) the Helmholtz function for the
electric field is real whereas the Schr\"odinger wave function is
complex. It would therefore be preferable to use a reliable and
consistent quantum mechanical formalism for photons.

Fortunately, such a formalism exists \cite{Ghose}, and is based on
the classic works of Kemmer \cite{Kemmer} and Harish-Chandra
\cite{HC}.  In this formalism, the wave function for the photon,
which obeys a first-order equation similar to the Dirac equation,
is a ten component column whose first six elements (the electric
and magnetic field strengths) are real functions and the last four
are zero, and there is a conserved four-vector current associated
with energy flow (not charge flow as in the familiar case of
charged particles with a complex wave function) whose time
component is positive definite and can be interpreted as a
probability density. The phase of such a wave function is
obviously not expressible as a multiplicative exponential factor
but is rather given in the same way as in classical
electrodynamics through an additive term in the sinusoidal
function for the fields. The signal velocity can be calculated in
this formalism unambiguously from the energy flux vector which
turns out to be proportional to the Poynting vector, as one would
expect.

It is the purpose of this paper to show, using this formalism,
that Einstein causal electromagnetic signals {\it can} indeed
travel faster than $c$ while tunneling through a photonic band gap
provided that the dielectric in the gap is inhomogeneous and
(practically) non-dispersive. The same result will be shown to
hold for classical light.

\section{The Tunneling Solution in Electrodynamics}

Let us consider the usual tunneling problem with a thin
non-magnetic, practically non-absorptive material with a band gap
around the frequency $\omega$, extending from $x = 0$ to $x = d$
and the signal incident normally on it so that there is no
dispersion. It is essentially a two-dimensional problem (in the
$x-y$ plane) expressible in terms of a single component of the
electric or magnetic field \cite{BW}. We will consider the case of
electric polarization with $H_x = H_z = 0,\, E_x = E_y = 0$ and
$\mu = 0,\, \epsilon= \epsilon (x)$, $\epsilon_0 = 1$. The same
result will hold for magnetic polarization also. Then Maxwell's
equations can be written in the rest frame of the dielectric
material in the form
 \ben
\partial_y E_z = 0, \,\,\,\,\,\,\,  \partial_x E_z &=&
\frac{1}{c} \partial_t H_y
\label{eq:1}\\
\partial_z H_y = 0, \,\,\,\,\,\,\, \partial_x H_y &=&
\frac{\epsilon(x)}{c}
\partial_t E_z
\label{eq:2}\\
\partial^2_{x} E_z - \frac{\epsilon (x)}{c^2} \partial^2_t E_z &=& 0
\label{eq:3} \een
\be
\partial^2_x H_y + \partial^2_y H_y + (\partial_x {\rm ln \epsilon(x)}) \partial_x H_y -
\frac{\epsilon (x)}{c^2}\partial^2_t H_y = 0 \label{eq:4} \ee

Let us first assume that the time variation of the electric and
magnetic fields is given by ${\rm exp}(\pm i \omega t)$, and use
the ansatz $E_z (x, y) = Y(x) U(y)$. Then it is easy to show that

\ben
U(y) = \beta e^{\pm i \frac{\omega}{c} \alpha y}
\een
where $\beta$ and $\alpha$ are constants. It follows from (\ref{eq:1})
that $\alpha = 0$, and so we have

\ben E_z = \beta Y(x) e^{\pm i \omega t} \label{eq:5}\\ H_y =
\frac {\mp i c \beta}{\omega} \frac{d Y(x)}{d x} e^{\pm i \omega
t} \label{eq:6} \een This shows that the magnetic field $H_y$ is
completely determined by the electric field $E_z$. It also follows
from (\ref{eq:1}) and (\ref{eq:2}) that

\ben
\frac{d Y(x)}{d x} = - \frac{\omega^2}{c^2} \int \epsilon (x) Y(x) d x
\een
or,

\ben
\frac{d^2 Y(x)}{d x^2} + \frac{\omega^2}{c^2} \epsilon (x) Y
(x) = 0 \label{eq:7}
\een
An approximate solution to this equation (\ref{eq:7}) is given by

\be
Y(x) \approx [k(x)]^{-\frac{1}{2}} \big[ c_1 \exp [-i \int_0^x
k(x) dx] + c_2 \exp[ i \int k(x) dx]\big] \ee where $k =
\sqrt{\epsilon(x)}\omega/c$, $c_1$ and $c_2$ are arbitrary
constants, and we have assumed that the change in $\epsilon(x)$
over one wavelength ($2 \pi/k$) is sufficiently small compared to
$|\epsilon(x)|$ (WKB approximation). This gives the usual
oscillating solution of $E_z(x, t)$:

\be
E_z(x, t) \approx [k(x)]^{-\frac{1}{2}} \big[ c_1 \exp [- i(
\int_0^x k(x) dx - \omega t)] + c_2 \exp [i (\int_0^x k(x) dx -
\omega t)]\big] \label{eq:osc} \ee Since the dielectric has a band
gap around the frequency $\omega$, these oscillating solutions
cannot propagate through it. One has to look for exponential or
tunneling solutions. In the case of the non-relativistic
Schr\"{o}dinger equation such solutions are obtained when the
function corresponding to $\epsilon (x)$, namely, $[E - V(x)]$,
becomes negative. This is not possible in electrodynamics because
$\epsilon (x)$ is never negative. However, it is significant that
a general tunneling solution can still be found, and is given by

\be
E^d_z(x, t) \approx [\kappa(x)]^{-\frac{1}{2}} \big[ c_1 \exp [ -
\int_0^x \kappa(x) dx +  \omega t ] + c_2 \exp [\int_0^x \kappa(x)
dx - \omega t]\big] \label{eq:tun}) \ee with $\kappa (x) = \omega
\sqrt{\epsilon (- i x)}/c$ a real, positive function\cite{Merz}.
This is clearly a solution of the wave equation

\be
\partial^2_{x} E^d_z - \frac{\epsilon (- i x)}{c^2} \partial^2_t E^d_z = 0
\label{eq:10} \ee which is Lorentz invariant as long as $\epsilon
(- i x)$ is a real, positive Lorentz scalar function. That is
guaranteed if $\epsilon (x, t)$ is a real, positive definite
function of the Lorentz invariant variable $(x^2 - c^2 t^2)$ in an
arbitrary inertial frame. We will therefore restrict our
discussions to such cases only.

Notice that the tunneling solution (\ref{eq:tun}) is a mapping of
the oscillating solution (\ref{eq:osc}) by

\ben x \rightarrow - i x ,\, \,\,\,\,\,\, t \rightarrow - i t
\label{eq:8} \een Maxwell's equations in vacuo are invariant under
this mapping. Maxwell's equations in an inhomogeneous dielectric
[equations (1) - (4)] are also invariant provided $\epsilon (- i
x)= \epsilon (x)$. But that is certainly not the most general
case. Assuming that $\epsilon (x)$ is an analytic function, one
can express it as a Taylor series around $x = 0$:

\be
\epsilon (x) = {\rm \epsilon}_0 + \sum_n a_n x^n\ee with the sum
positive definite \cite{Somm}. Thus $\epsilon (- i x)$ will be
complex in general. But, since ${\rm Im} \sqrt{\epsilon ( - i x)}$
will give rise to oscillating terms in (\ref{eq:tun}), and since
the material is assumed to have a band gap around $\omega$, it
must vanish. Maxwell's equations then get mapped on to equations,
such as equation (\ref{eq:10}), that are still Lorentz invariant
and therefore acceptable. It is clear from equation (\ref{eq:10})
that the propagation will be superluminal provided $\epsilon (
-ix) < \epsilon_0 (=1)$. This is possible, for example, if the
dielectric function $\epsilon ( -ix) = (1 + \sum_n a_n x^n) < 1$
with $n$ such that ${\rm Im} \sqrt{\epsilon ( - i x)} = 0$ and
$\sum_n a_n x^n < 0$.

An immediate consequence of the mapping (\ref{eq:8}) is that
time-like intervals are mapped on to space-like intervals $\,(c^2
t^2 - x^2)\rightarrow (x^2 - c^2 t^2)\,$. Consequently, if
$\epsilon ( -ix) < 1$, {\it all causally related events get
connected by superluminal signals}. Conversely, it is
straightforward to see that superluminal signals ($v > c$) imply
the mapping (\ref{eq:8}), because

\ben x' &=& (x - v t)/\sqrt{1 - v^2/c^2} = - i(x - v
t)/\sqrt{v^2/c^2 - 1}\nonumber\\ t' &=& (t - v x/c^2)/\sqrt{1 -
v^2/c^2} = - i(t - v x/c^2)/\sqrt{v^2/c^2 - 1}\een This is
remarkable and important for the interpretation of the experiments
showing superluminal tunneling--- they  do not contradict Lorentz
invariance and causality.

It is instructive to look at the difference between superluminal
optical tunneling and tunneling of massive particles. While
tunneling, the energy and momentum of massive relativistic
particles are imaginary, as one can easily verify by applying the
energy and momentum operators on their wavefunction. Thus, the
relativistic relation $E^2 = p^2 c^2 + m_0^2 c^4$ gets mapped on
to $E^2 = p^2 c^2 - m_0^2 c^4$, implying tachyons. This does not
happen for massless bosons. Nevertheless, as we have seen above,
tunneling solutions in electrodynamics are also superluminal.

It is often asserted that according to the special principle of
relativity the maximum velocity that a physical signal can have is
the velocity of light $c$ in vacuum. If that is correct, then the
special relativity principle would rule out the possibility of
dielectric materials of the kind discussed above. That would imply
that somehow only dielectrics with the property $\epsilon (-i x) =
\epsilon (x)$ can exist physically. Whereas that is not
impossible, we find it hard to believe that such a demonstration
can indeed be given. On the other hand, if one restricts oneself
to the assumptions actually made by Einstein, namely the postulate
of relativity of uniform motion coupled with the postulate that
the velocity of light is independent of the motion of the light
source, one need only insist on Lorentz invariance as a necessary
condition for a physical law \cite{Pauli}. That would leave open
the possibility of dielectrics of the kind that would make
superluminal yet causal signals possible in tunneling modes.

Interestingly, the dielectrics chosen in the tunneling experiments
\cite{Steinberg} all had variable layers of dielectrics and were
practically dispersion free. Now, it is well-known that causality
and dispersion relations are intimately related \cite{toll}. It
follows from these dispersion relations that the real part of the
refractive index $n$ must vanish for a purely non-dispersive
material. Hence the velocity of propagation $c/n$ of light through
such a material has no upper limit. The problem is to produce such
materials. The trick is to prepare a medium in such a way that it
is inhomogeneous with alternative thin layers of high and low
refractive indices $n_i$ that are all greater than unity ($n_i >
1$) so that it acquires a photonic band gap. Then the evanescent
wave sees a refractive index $< 1$, as we have seen, and so
propagates superluminally without changing shape.

\section{Quantum Mechanical Formulation of Optical Tunneling}

We will now show how to give a purely quantum mechanical
formulation of this superluminal tunneling behaviour. For this we
need to use a consistent quantum mechanical formulation of
massless electrodynamics using the Kemmer--Harish-Chandra
formalism \cite{Ghose}, outlined in the Appendices. It is clear
from this formalism that the classical Maxwell fields are
components of a ten-component quantum mechanical wavefunction with
constraints that reduce the degrees of freedom to two. For the
tunneling problem, the number of degrees of freedom is further
reduced to one, as we have already seen. Let the incident finite
duration signal be represented by the electric fields (components
of the ten dimensional unnormalized photon wavefunction $\gamma
\psi$, vide Appendix A) \ben E_z^i &=& \int d k A(k)\, \cos{(kx
-\omega t - \phi)} - \nonumber\\ &\sqrt{R}&\int d k A(k)\,
\cos{(kx + \omega t + \phi)}\, \,\,\,\, {\rm for}\,\, x \leq 0\\
E_z^d &=& \theta(t) \frac{1}{\sqrt{\kappa(x)}} C \exp{[- \int_0^x
\kappa(x) dx + \omega_0 t]}\,\,\,\, {\rm for}\,\, 0\leq x\leq d
\een \be E_z^f = \theta(t - \tau )\sqrt{T} \int d k A(k)\,
\cos{[k(x - d) -\omega (t - \tau) + \chi]}\:\:\: {\rm for}\:\:\: x
\geq d \ee where $A(k) = (1/\sqrt{2 \pi \sigma^2})\, \exp{[ - (k -
k_0)^2/2 \sigma^2]}$ is real and $\int_{- \infty}^{\infty} A(k) d
k = 1$, $\int_{- \infty }^{\infty} k A(k) d k = k_0$. $R$ and $T$
are the reflection and transmission coefficients, $k = \omega/c$,
$\kappa(x) = k_0 \sqrt{\epsilon(-i x)}$, $\tau$ is the tunneling
or dwell time and $\theta(t)$ is the step function. (Note that
there is no term representing a reflected wavefunction within the
tunneling region because we are not considering a steady state
situation or times $t > \tau$.) Accordingly, the dielectric medium
is at rest (in the sense of being free of any disturbance) before
$t=0$ and there is no emerging signal at $x = d$ before $t =
\tau$. By matching the wavefunctions smoothly at the boundary
$x=0,\, t =0$, we get

\ben C &=& \sqrt{\kappa(0)}(1 - \sqrt{R})\, \cos{\phi} \\ \tan{
\phi} &=& \frac{\kappa(0)}{k_0} = 1\een Hence

\be
E_z^d = \theta(t)\frac{ \sqrt{\kappa(0)}(1 - \sqrt{R})\, \cos{
\phi}}{\sqrt{\kappa (x)}}\, {\rm exp} [-\int_0^x \kappa(x) dx +
\omega_0 t ]\ee The magnetic field in the tunneling region is
determined by the analog of (\ref{eq:6}) for the tunneling case
and is given by

\be
H_y^d = \theta(t)\frac{c}{\omega_0} \partial_x E_z^d\ee Therefore
we have (in the WKB approximation)

\be
H_y^d = -\theta(t)\sqrt{\kappa(0)}(1 -
\sqrt{R})\sqrt{\kappa(x)}\,\cos{\phi} \frac{c}{\omega_0}\, {\rm
exp} [- \int_0^x \kappa(x) dx + \omega_0 t ]\ee

Matching the wavefunctions at the other boundary $x = d,\, t =
\tau$ gives

\be
\sqrt{T} = \frac{\sqrt{\kappa(0)}(1 -\sqrt{R})\, \cos{\phi}
}{\sqrt{\kappa(d)}}\,\sec{\chi} \:\: {\rm exp}\,[- \int_0^d
\kappa(x) dx + \omega_0 \tau]\ee Further, matching the derivatives
of the wavefunctions at this boundary, one has

\be
\tan{\chi} = \frac{\kappa(d)}{k_0}\label{eq:tau} \ee

The velocity operator in this formalism is the $10 \times 10$
matrix $v \tilde{\beta}_x = (c/\sqrt{\epsilon(-ix)})\\ ( \beta_0
\beta_x - \beta_x \beta_0)$. Thus the Poynting vector can now be
calculated, and is given by (see Appendix A)

\ben S_x^d &=& m_0 c^3\psi^{T}\gamma \tilde{\beta_x} \gamma\psi =
- c E_z^d H_y^d\nonumber\\&=& \theta(t) \kappa(0)(1 -
\sqrt{R})^2\, {\rm cos^2 \phi} \,\frac{c^2}{2 \omega_0}\,{\rm
exp}\, [ - 2\, (\int_0^x \kappa(x) dx - \omega_0 t)]\een The
energy density is given by (see Appendix A)

\ben {\cal{E}}^{d} &=& \frac{1}{2}\psi^{T}\gamma\psi =
\frac{1}{2}[\epsilon (-ix) E_z^{d 2} + H_y^{d 2}]\nonumber\\ &=&
\theta(t)\, \kappa(0)(1 - \sqrt{R})^2\, {\rm cos^2 \phi}
\frac{c^2}{2 \omega_0^2}\kappa(x)\, {\rm exp}\, [ - 2\, (\int_0^x
\kappa(x) dx - \omega_0 t)]\een One can therefore calculate the
velocity of energy transport

\be
v_x^d = \frac{S_x}{{\cal{E}}^d} =
\frac{c}{\sqrt{\epsilon(-ix)}}\ee It follows from this that the
tunneling time is given by

\be \tau = \int_0^d \frac{d x}{v_x^d }\ee which implies

\be
\int_0^d \kappa(x) dx - \omega_0 \tau = 0\ee In a hypothetical
model in which $\sqrt{\epsilon ( -i x)} = 1 - a x^2$,

\be
\tau = \frac{d}{c} - \frac{ad^3}{3 c}\ee which is always less than
the time for passage through vacuum. This superluminal effect will
be further accentuated if one includes higher order terms in $x$
in the expansion of $\sqrt{\epsilon ( -i x)}$ because of the
condition $\sum_n a_n x^n < 0$ stated above.

If one uses the de Broglie-Bohm guidance condition $v_x^d = d x/d
t$, one again obtains the same result for $\tau$. These results
confirm that the energy and so the physical signal indeed
propagates superluminally while tunneling.

\section{Conclusions}

In conclusion we would like to emphasize precisely the significant
new result that we have obtained. Since there has been much
discussion and some controversy in the literature regarding
superluminal effects and their causality, let us summarize the
situation as we see it.

The materials used for observing superluminal effects have been
generally termed ``ultrarefractive'' \cite{ultraref}. Near the
edges of a transmission gap the effective permitivity can become
close to zero. Consequently, surprising effects can be observed on
light transmitted and reflected by such materials, such as
superluminal velocities as well as enlargement and splitting of
the transmitted beam.

In one type of process the effects are results of anomalous
dispersion, i.e., anomalous variation of the permitivity with
wavelength. Although the 1914 analysis of Sommerfeld and Brillouin
clearly established that superluminality in such cases cannot be
Einstein causal and is only apparent, it has recently been argued
that this need not be the case for physical signals that are of
finite duration and extent because a responsive signal travelling
backward in time in such a case cannot arrive before the primary
signal is generated, thus preserving the causality principle
\cite{nimtz}. Our paper does not deal with this type of phenomena.

The second type of process involves tunneling in one (or two)
dimensions through a narrow band gap, and it is only this type of
phenomena ($1$D tunneling) that we have addressed. The theoretical
discussions of such phenomena have so far been based purely on an
analogy between the non-relativistic Schr\"odinger equation and
the Helmholtz equation leading to an effective refractive index
$n(x,y,z) = \{2m [E - V(x,y,z)]\}^{1/2} c/\hbar \omega$ which is
imaginary in any region where $E < V$ \cite{Steinberg,Rev1}. This
mechanism is, in reality, not applicable to photons, as we have
mentioned earlier and as Chiao and Steinberg admit in their review
article \cite{Rev1}. To take a definite stand on an issue such as
superluminal propagation and causality, analogies are not reliable
in our opinion, and one must use a proper theory, namely a
consistent relativistic quantum mechanical formalism for photons
\cite{Ghose}. We have used this formalism to carry out explicit
calculations for the tunneling of a finite width photon
wave-packet incident normally on a $1$D photonic barrier. (Note
that in this sense also our result is new because total internal
reflection in optics occurs only for {\it non-zero} critical
angles of incidence.) Our analysis clearly shows that genuine
Einstein causal superluminal propagation can occur only if the
tunneling medium is inhomogeneous on the scale of the wavelength
and ${\rm Im}\, \epsilon (-ix) = 0$. This follows simply and very
generally from the fact that points on the light cone remain on
the light cone under the mapping (\ref{eq:8}) which takes
propagating solutions to tunneling solutions. Therefore, the only
way to get genuine superluminal signals is to have an
inhomogeneous dielectric function $\epsilon (x) > 1$ that is
mapped to $\epsilon (-ix) < 1$ with ${\rm Im}\, \epsilon (-ix) =
0$ to ensure Lorentz invariance of the wave equation
(\ref{eq:10}). This argument obviously holds for both classical
and quantum light, and is consistent with dispersion relations and
causality \cite{toll}.

Such materials have been used in actual experiments
\cite{Rev1,Rev2}. They involve tunneling at near normal incidence
through band gaps excited in periodic dielectric structures. These
band gaps arise from Bragg reflections from the periodic
structure, leading to an evanescent decay of the wave amplitude
when the frequency is within the forbidden band gap at the first
Brillouin zone. It should be noted that such periodic structures
are {\it non-dispersive} so that the tunneling wave-packets that
are tuned to midgap remain essentially undistorted upon
transmission through the barrier, though much attenuated in
amplitude \cite{Rev1}.

\section{Acknowledgement}
The authors thank DST, Govt. of India for financial support to
carry out this work.

\section{Appendix A}

Until recently, no consistent quantum mechanical formalism existed
for relativistic bosons below the threshold for pair production
and annihilation. Relativistic quantum mechanics can only be
consistently formulated provided there exists a conserved
four-vector current whose time component, to be identified with
the probability density, is positive definite. Unfortunately, the
conserved charge vector current for relativistic spin $0$ and spin
$1$ bosons does not have this property. Moreover, the charge
current vanishes for neutral particles like the photon. However,
it has now been shown \cite{Ghose} that a conserved four-vector
current with a positive definite time component does exist for
relativistic bosons, and is associated, not with the charge
current but, with the flow of energy. This formulation is based on
the first-order Kemmer equation \cite{Kemmer}

\be
(\,i\,\hbar\,\beta_\mu\,\partial^\mu + m_0\,c\,)\,\psi = 0\,
\label{eq:1A} \ee where the matrices $\beta$ satisfy the algebra

\be
\beta_{\mu}\,\beta_{\nu}\,\beta_{\lambda} +
\beta_{\lambda}\,\beta_{\nu}\, \beta{\mu} = \beta_{\mu}\,g_{\nu
\lambda} + \beta_{\lambda}\,g_{\nu \mu}\,. \label{eq:2A} \ee The
$5\times 5$ dimensional representation of these matrices describes
spin 0 bosons and the $10 \times 10$ dimensional representation
describes spin 1 bosons. Multiplying (\ref{eq:1A}) by $\beta_0$,
one obtains the Schr\"{o}dinger form of the equation

\be
i\,\hbar\,\frac{\partial \psi}{d t} = [\,- i\,\hbar\,c\,
\tilde{\beta}_i\,
\partial_i - m_0\,c^2\,\beta_0\,]\,\psi
\label{eq:3A} \ee where $\tilde{\beta}_i \equiv \beta_0\,\beta_i -
\beta_i\,\beta_0$. Multiplying (\ref{eq:1A}) by $1- \beta_0^2$,
one obtains the first class constraint

\be
i\,\hbar\,\beta_i\,\beta_0^2\,\partial_i\,\psi = -m_0\,c\,(\,1 -
\beta_0^2\,) \,\psi. \label{eq:4A} \ee It implies the conditions
${\rm div} \vec{D} = - (m_0^2 c/\hbar) A_0$ and $\vec{B}= {\rm
curl} \vec{A}$ if one takes

\be
\psi^T = (1/\sqrt{m_0 c^2}) \\( -D_x, -D_y, -D_z, B_x, B_y, B_z,
-m_0 A_x, -m_0 A_y, -m_0 A_z, m A_0))\ee The reader is referred to
Ref. \cite{Ghose} for further discussions regarding the
significance of this constraint.

If one multiplies equation (\ref{eq:3A}) by $\psi^{\dagger}$ from
the left, its hermitian conjugate by $\psi$ from the right and
adds the resultant equations, one obtains the continuity equation

\be
\frac{\partial\,( \psi^{\dagger}\,\psi )}{\partial t} +
\partial_i\, \psi^{\dagger}\,\tilde{\beta}_i\,\psi = 0\,.
\label{eq:5A} \ee This can be written in the form

\be
\partial^\mu\,\Theta_{\mu 0} = 0
\ee where

\be
\Theta_{\mu \nu} = - m_0 c^2 \bar{\psi}(\beta_{\mu} \beta_{\nu} +
\beta_{\nu} \beta_{\mu} - g_{\mu \nu})\psi\ee (with $\bar{\psi} =
\psi^{\dagger} \eta_0$, $\eta_0 = 2 \beta_0^2 - 1, \eta_0^2 = 1$)
is the symmetric energy-momentum tensor, and \be \Theta_{0 0} = -
m_0 c^2 \psi^{\dagger}\,\psi < 0\ee Thus, it is possible to define
a wave function $\phi = \sqrt{m_0 c^2/E}\,\psi$ (with $E =-
\int\,\Theta_{0 0} \,dV$ ) such that $\phi^{\dagger}\,\phi$ is
non-negative and normalized and can be interpreted as a
probability density. The conserved probability current density is
$s_\mu = - \Theta_{\mu 0}/E = (\,\phi^{\dagger}\,\phi, -
\phi^{\dagger}\,\tilde{\beta}_i\,\phi )$.

Notice that according to the equation of motion (\ref{eq:3A}), the
velocity operator for massive bosons is $c\,\tilde{\beta}_i$.

The theory of massless spin 0 and spin 1 bosons cannot be obtained
simply by taking the limit $m_0$ going to zero because of the
$1/\sqrt{m_0}$ factor in $\psi$. One has to start with the
equation \cite{HC}

\be
i\,\hbar\,\beta_\mu \partial^\mu\,\psi + m_0\,c\,\gamma\,\psi = 0
\label{eq:8A} \ee where $\gamma$ is a matrix that satisfies the
following conditions:

\ben \gamma^2 &=& \gamma\,\\ \gamma\,\beta_\mu + \beta_\mu\,\gamma
&=& \beta_\mu\,. \label{eq:9} \een This equation can be derived
from the gauge invariant Lagrangian density

\be
{\cal{L}} = - \frac{i \hbar}{2}[\partial^{\mu} \bar{\psi} \gamma
\beta_{\mu} \psi - \bar{\psi} \beta_{\mu} \gamma \partial^{\mu}
\psi] + \frac{m_0 c}{2} \bar{\psi} \gamma \psi \ee

Multiplying (\ref{eq:8A}) from the left by $1 - \gamma$, one
obtains
\be
\beta_\mu\,\partial^\mu\, (\,\gamma\,\psi\,) = 0\,. \label{eq:10A}
\ee Multiplying (\ref{eq:8A}) from the left by
$\partial_{\lambda}\, \beta^{\lambda}\,\beta^{\nu}$, one also
obtains

\be
\partial^{\lambda}\,\beta_{\lambda}\,\beta_\nu\,(\,\gamma\,\psi\,) =
\partial_\nu\, (\,\gamma\,\psi\,)\,.
\label{eq:11A} \ee It follows from (\ref{eq:10A}) and
(\ref{eq:11A}) that

\be
\Box\,\, (\,\gamma\,\psi\,) = 0 \label{eq:12A} \ee which shows
that $\gamma\,\psi$ describes massless bosons.

The Schr\"{o}dinger form of the equation

\be
i\,\hbar\,\frac{\partial\, (\,\gamma\,\psi\,)}{d t} = -
i\,\hbar\,c \tilde{\beta}_i\,\partial_i\, (\gamma\,\psi)
\label{eq:13A} \ee and the associated first class constraint

\be
i\,\hbar\,\beta_i\,\beta_0^2\,\,\partial_i\,\psi + m_0\,c\,(\,1 -
\beta_0^2\,)\,\gamma\,\psi = 0 \label{eq:14A} \ee follow by
multiplying (\ref{eq:8A}) by $\beta_0$ and $1 - \beta_0^2$
respectively. Equation (\ref{eq:13A}) implies the Maxwell
equations ${\rm curl} \vec{E} = -(\mu/c)
\partial_t \vec{H}$ and ${\rm curl} \vec{H} = (\epsilon/c)
\partial_t \vec{E}$ if

\be
\gamma \psi^T = (1/\sqrt{m_0 c^2}) \\ ( -D_x, -D_y, -D_z, B_x,
B_y, B_z, 0, 0, 0, 0)\nonumber\ee The constraint (\ref{eq:14A})
implies the relations ${\rm div} \vec{E} = 0$ and $\vec{B} = {\rm
curl}\vec{A}$. The symmetrical energy-momentum tensor is
\be
\Theta_{\mu \nu} = -\frac{ m_0 c^2}{2} \bar{\psi}(\beta_{\mu}
\beta_{\nu} + \beta_{\nu} \beta_{\mu} - g_{\mu \nu})\gamma\psi\ee
and so the energy density

\ben {\cal{E}} = -\Theta_{0 0} = \frac{m_0 c^2}{2}
\psi^{\dagger}\,\gamma \psi =  \frac{1}{2}[\vec{E}. \vec{E} +
\vec{B}. \vec{B} ]\een is positive definite. The rest of the
arguments are analogous to the massive case.

The Bohmian 3-velocity $v_i$ for massless bosons can be defined by

\be
v_i = c \frac{\psi^T \gamma \tilde{\beta}_i \gamma\psi}{\psi^T
\gamma \psi}\ee Notice that in relativistic quantum mechanics the
Bohmian velocity is not defined through the gradient of the phase
as in non-relativistic quantum mechanics but in terms of the
energy flux current.

Neutral massless vector bosons are very special in quantum
mechanics. Their wave function is real, and so their charge
current $j_\mu = \psi^{T}\,\beta_\mu\,\gamma\psi$ vanishes.
However, their probability current density $s_\mu$ does not
vanish. Furthermore, the Poynting vector turns out to be given by

\be
S_i = m_0 c^3 \psi^T \gamma\tilde{\beta}_i\gamma \psi = c [\vec{E}
\times \vec{H} ]_i\ee

One might wonder about the significance of the mass parameter
$m_0$ for massless electrodynamics. It is necessary for a
consistent quantum mechanical formalism for dimensional reasons
and drops out of all physical results because of the operator
$\gamma$. It can be altogether eliminated in favour of the
intrinsic parameters in the theory, namely $c$, $\hbar$, the
frequency $\omega$ and the spin multiplicity $s$.

The representations of the Kemmer-Duffin-Petiau $\beta$ matrices
used in this paper are given in Appendix B.

\section{Appendix B}\vskip 0.3in{\tiny$i\beta_1 = \pmatrix{ 0 & 0 &
0 & . & 0 & 0 & 0 & . & 0 & 0 & 0 & . & -1\cr 0 & 0 & 0 & . & 0 &
0 & 0 & . & 0 & 0 & 0 & . & 0\cr 0 & 0 & 0 & . & 0 & 0 & 0 & . & 0
& 0 & 0 & . & 0\cr . & . & . & . & . & . & . & . & . & . & . & . &
. & .\cr 0 & 0 & 0 & . & 0 & 0 & 0 . & & 0 & 0 & 0 & . & 0\cr 0 &
0 & 0 & . & 0 & 0 & 0 & . & 0 & 0 & -1 & . & 0\cr 0 & 0 & 0 & . &
0 & 0 & 0 & . & 0 & 1 & 0 & . & 0\cr . & . & . & . & . & . & . & .
& . & . & . & . & . & .\cr 0 & 0 & 0 & . & 0 & 0 & 0 & . & 0 & 0 &
0 & . & 0\cr 0 & 0 & 0 & . & 0 & 0 & 1 & . & 0 & 0 & 0 & . & 0\cr
0 & 0 & 0 & . & 0 & -1 & 0 & . & 0 & 0 & 0 & . & 0\cr . & . & . &
. & . & . & . & . & . & . & . & . & . & . \cr -1 & 0 & 0 & . & 0 &
0 & 0 & . & 0 & 0 & 0 & . & 0}\,\,i\beta_2 = \pmatrix{ 0 & 0 & 0 &
. & 0 & 0 & 0 & . & 0 & 0 & 0 & . & 0\cr 0 & 0 & 0 & . & 0 & 0 & 0
& . & 0 & 0 & 0 & . & -1\cr 0 & 0 & 0 & . & 0 & 0 & 0 & . & 0 & 0
& 0 & . & 0\cr . & . & . & . & . & . & . & . & . & . & . & . & . &
. \cr 0 & 0 & 0 & . & 0 & 0 & 0 & . & 0 & 0 & 1 & . & 0\cr 0 & 0 &
0 & . & 0 & 0 & 0 & . & 0 & 0 & 0 & . & 0\cr 0 & 0 & 0 & . & 0 & 0
& 0 & . & -1 & 0 & 0 & . & 0\cr . & . & . & . & . & . & . & . & .
& . & . & . & . & . \cr 0 & 0 & 0 & . & 0 & 0 & -1 & . & 0 & 0 & 0
& . & 0\cr 0 & 0 & 0 & . & 0 & 0 & 0 & . & 0 & 0 & 0 & . & 0\cr 0
& 0 & 0 & . & 1 & 0 & 0 & . & 0 & 0 & 0 & . & 0\cr . & . & . & . &
. & . & . & . & . & . & . & . & . & . \cr 0 & -1 & 0 & . & 0 & 0 &
0 & . & 0 & 0 & 0 & . & 0}$ \vskip 0.5 in \noindent $i\beta_3 =
\pmatrix{ 0 & 0 & 0 & . & 0 & 0 & 0 & . & 0 & 0 & 0 & . & 0\cr 0 &
0 & 0 & . & 0 & 0 & 0 & . & 0 & 0 & 0 & . & 0\cr 0 & 0 & 0 & . & 0
& 0 & 0 & . & 0 & 0 & 0 & . & -1\cr . & . & . & . & . & . & . & .
& . & . & . & . & . & .\cr 0 & 0 & 0 & . & 0 & 0 & 0 & . &  0 & -1
& 0 & . & 0\cr 0 & 0 & 0 & . & 0 & 0 & 0 & . & 1 & 0 & 0 & . &
0\cr 0 & 0 & 0 & . & 0 & 0 & 0 & . & 0 & 0 & 0 & . & 0\cr . & . &
. & . & . & . & . & . & . & . & . & . & . & .\cr 0 & 0 & 0 & . & 0
& 1 & 0 & . & 0 & 0 & 0 & . & 0\cr 0 & 0 & 0 & . & -1 & 0 & 0 & .
& 0 & 0 & 0 & . & 0\cr 0 & 0 & 0 & . & 0 & 0 & 0 & . & 0 & 0 & 0 &
. & 0\cr . & . & . & . & . & . & . & . & . & . & . & . & . & . \cr
0 & 0 & -1 & . & 0 & 0 & 0 & . & 0 & 0 & 0 & . & 0}\,\,\beta_0 =
\pmatrix{ 0 & 0 & 0 & . & 0 & 0 & 0 & . & -i & 0 & 0 & . & 0\cr 0
& 0 & 0 & . & 0 & 0 & 0 & . & 0 & -i & 0 & . & 0\cr 0 & 0 & 0 & .
& 0 & 0 & 0 & . & 0 & 0 & -i & . & 0\cr . & . & . & . & . & . & .
& . & . & . & . & . & . & .\cr 0 & 0 & 0 & . & 0 & 0 & 0 & . & 0 &
0 & 0 & . & 0\cr 0 & 0 & 0 & . & 0 & 0 & 0 & . & 0 & 0 & 0 & . &
0\cr 0 & 0 & 0 & . & 0 & 0 & 0 & . & 0 & 0 & 0 & . & 0\cr . & . &
. & . & . & . & . & . & . & . & . & . & . & .\cr i & 0 & 0 & . & 0
& 0 & 0 & . & 0 & 0 & 0 & . & 0\cr 0 & i & 0 & . & 0 & 0 & 0 & . &
0 & 0 & 0 & . & 0\cr 0 & 0 & i & . & 0 & 0 & 0 & . & 0 & 0 & 0 & .
& 0\cr . & . & . & . & . & . & . & . & . & . & . & . & . & .\cr 0
& 0 & 0 & . & 0 & 0 & 0 & . & 0 & 0 & 0 & . & 0}$}

\end{document}